# ECONOMIC PROSPECTS OF THE RUSSIAN-CHINESE PARTNERSHIP IN THE LOGISTICS PROJECTS OF THE EURASIAN ECONOMIC UNION AND THE SILK ROAD ECONOMIC BELT: A SCIENTIFIC LITERATURE REVIEW

**Elena Rudakova,**

*PhD in Law, Associate Professor*
*Law Institute of the Russian University of Transport, Russia, Moscow*

**Alla Pavlova,**

*PhD in Law, Associate Professor*
*Law Institute of the Russian University of Transport, Russia, Moscow*

**Oleg Antonov,**

*PhD in Law, Associate Professor*
*Law Institute of the Russian University of Transport, Russia, Moscow*

**Kira Kuntsevich,**

*Lecturer, RANEPA, Russia, Moscow*

**Yue Yang,**

*Associate Professor, Hebi University, China*

**Abstract**

*The authors of the article have reviewed the scientific literature on the development of the Russian-Chinese cooperation in the field of combining economic and logistics projects of the Eurasian Economic Union and the Silk Road Economic Belt. The opinions of not only Russian, but also Chinese experts on these projects are indicated, which provides the expansion of the vision of the concept of the New Silk Road in both countries.*

## 1. Introduction

The topic of the research is extremely important due to the fact that the relations between China and Russia at the present stage have a significant impact on the development of the entire system of international relations. Both China and

Russia are permanent members of the UN Security Council and have great economic potential, and are also included in the list of the ten leading countries in the world in terms of GDP, have solid nuclear missile weapons and their means of delivery, have great political weight in the international arena in the system of modern international relations. Russia and China have established 'strategic partnership relations turned into the twenty-first century', concluded an agreement on 'Good-neighborliness, Friendship and Cooperation', and thereby created a strong legal basis for the development of stable relations between these states.

## 2. Main part

China is becoming an increasingly powerful geopolitical player in the political arena and is taking a more active part in international politics, including in such organizations as APEC, SCO, the UN, BRICS, etc. Such politics allows China to realize its national economic, political, and cultural interests.

The Russian Federation, just like China, plays a significant role in the world. In fact, still a young player in the geopolitical arena of the world, Russia has been able to achieve considerable success in strengthening and defending its influence in the world, including through the establishment of mutually beneficial Russian-Chinese relations.

The expansion of cooperation between Russia and China is explained not only by economic benefits and neighboring position. The two major player countries face common tasks, common challenges, and new opportunities are opening up. In our age of globalization, transparent economies and democratic values, they can be realized only by uniting.

Cooperation between the development of the Eurasian Economic Union and the construction of the Silk Road Economic Belt (see Fig. 1 and Fig.2) is of great importance for both countries, and makes a significant contribution to the process of economic globalization and political multipolarity.

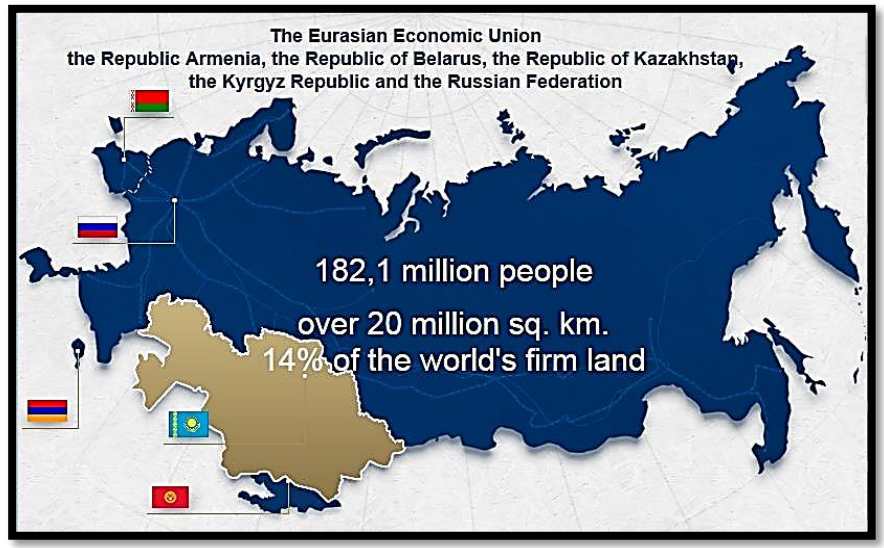

Fig.1. Eurasian Economic Union

*Source: Eurasian Economic Commission.*

The problem of studying the processes of conjugation of the Chinese Silk Road Economic Belt (SREB) and the Eurasian Economic Union( EAEU), as well as the opportunities for their interaction with the Shanghai Cooperation Organization (SCO), is rapidly becoming more active in the Russian, Chinese and Central Asian expert space.

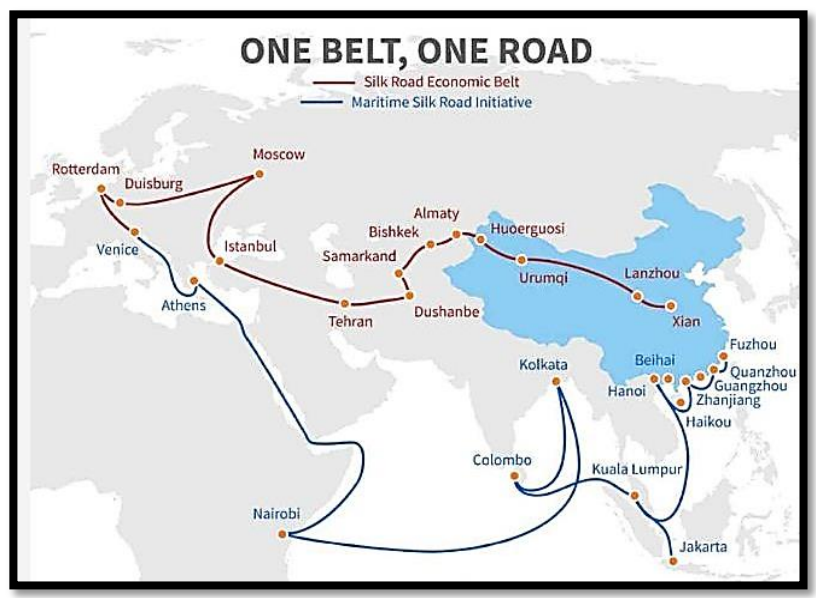

Fig.2. Silk Road Economic Belt

*Source: 'One Belt, One Road'. URL: https://eurasiainteraction.com/index.php/blog/175-one-belt-one-road-initiative-economic-strategic-prospects-for-southeast-asia*

Every year, dozens of different expert and analytical materials are published, written by both our sinologists and Chinese experts. The development of the research issues was carried out in the format of interpreting and commenting on scientific works that represent popular views on integration processes on the Eurasian continent in Russia, China and the West, and which can be a demonstration of the geopolitical picture in the Eurasian space.

The first group of works is devoted to the study of the Chinese concept of 'One Belt – One Road'. With the promotion of the concept of 'One Belt – One Road' in 2013, the dedicated topic is gradually attracting great attention from Russian scientific circles. The following Russian scientists worked on the analysis of the foreign economic concept adopted by China: Uyanaev S. V. [1; 2], Larin A. G. [3; 4], Lukin A.V. [5], Luzyanin S. G. [6], Remyga V. N. [7], Kuziev N. A. [8], whose works consider the main goals, contents, principles and methods of implementing the 'One Belt – One Road', explore the prospects and challenges for both the development of China and for Sino-Russian cooperation in the implementation of the strategy. These works helped to make a deep analysis of the characteristics of the Chinese project.

Chinese researchers are engaged in the most intensive development of this issue, among which Wu Jianming [9], Huang Yipin [10], Jin Lin [11], Su Ge [12], Shi Yan [13], Yuan Xintao [14] can be singled out. In their works, the researchers provide a detailed analysis of the political and economic aspects of the new Chinese strategy, demonstrate its essence and specifics. In particular, it is necessary to note Wu Jianmin's monograph 'The Belt and the Road' and Great-power Diplomacy with Chinese specifics', which evidently analyzes the motivation and specifics of the Chinese initiative under the changing external and internal environment.

The second group includes works devoted to the study of the Eurasian Economic Union. Since the Eurasian Economic Union is a necessary subject that directly shows Russia's interests in the regional arena.

The analysis of this group allows the author to reveal the interests of Russia and the possibilities of combining the two projects. Great attention is paid by Russian scientists such as Mansurov T. A. [16], Volodin V. M., Kaftulina Yu. A.,

Rusakova Yu. I. [17], Bekyashev K. A. and Moiseev E. G. [18]. Their works analyze the main provisions, the evolution of the Eurasian Economic Union, identify the prospects and opportunities of the EAEU on the path of the process of deep economic integration. It is necessary to note the works of N. A. Vasilyeva and M. L. Lagutina [19; 20], whose articles consider the issue of the integration paradigm on the example of the Eurasian Economic Union, comprehensively analyze the positions regarding the project of the Eurasian Economic Union.

The analysis of the issue of the Eurasian Economic Union is still contained in the works of Chinese scientists, such as Li Ziguo [21], Zhou Mi [22], Li Xin [23], Liu Qingcai and Zhi Zichao [24]. In their works, the authors investigated the situation of the EAEU member states, analyzed significant problems, risks of the EAEU impact, and identified possible solutions. Among them, it is worth noting the monograph by Fu Jingjun [25], where the scientist examines in detail the process of creating and developing the EAEU, highlights the approaches of the EAEU member states to the Chinese 'Silk Road Economic Belt', also analyzes how to implement the interface of the two projects. In addition, this topic includes the work of Chinese scientists Li Jianming and Li Yongquan [26], where the prerequisites, state, difficulties and prospects of the Eurasian Economic Union are revealed, as well as numerous factors affecting the possibility of combining this and the 'One Belt, One Road' project are investigated.

Russian scientists such as Vasiliev L. E. [27], Matveev V. A. [28], Glinkina S. P., Turaeva M. O. and Yakovlev A. A. are engaged in the study of the interrelationships between China and Central Asia [29]. The works of these scientists assess the current state of cooperation between China and Central Asia and characterize China's policy in Central Asia. Among Chinese researchers, Zhao Changqing [30], Yuan Shenyu and Wang Weimin [31] dealt in detail with issues related to China's relations with the countries of Central Asia.

The next group includes works devoted to the study of Russia's position on the Chinese initiative and Sino-Russian cooperation in the process of its implementation. Russian researchers such as Denisov I. E. [32], Vladimir

Berezhnoy [33], Morozov Yu. V. [34], Ostrovsky A.V. [35]. we worked on forming a position and analyzing Russia's place on the new Silk Road. Nevertheless, the problem of the Russian assessment of the concept of 'One Belt – One Road' occupies a significant place in the works of Chinese scientists-Wang Kei [36], Li Xiujiao [37], where they are devoted to considering the approaches of the Russian scientific community, identifying the reasons for changing the assessment with the international and domestic changing situation.

When analyzing the current Sino-Russian cooperation within the framework of 'One Belt – One Road', it is impossible to ignore Wang Qi's monograph [38], where the author comprehensively explains strategic cooperation in the fields of trade and economy, science and technology, military security, culture and education, considering the strategy for the development of bilateral relations, at the same time presents a proposal. It should be noted that Chinese researchers are engaged in the most intensive development of this issue, among which Zhang Jianping and Li Jing [39], Li Jianmin [40], Jiang Zhenjun [41] can be particularly distinguished. Their works touch upon the practical steps of interaction between China and Russia, problems and challenges in the process of cooperation in the joint creation of the 'One Belt, One Road'. In addition, the work of expert Zhao Huizhong is devoted to the issues of Sino-Russian cooperation within the framework of the 'One Belt, One Road' project, where the author interprets the importance of Russia in the Chinese project, as well as the importance of Sino – Russian joint cooperation in the implementation of the initiative, finds out the difficulties facing Beijing and Moscow, and presents objective proposals for their resolution [42].

3. **Conclusion**

The Sino-Russian comprehensive strategic partnership meets the fundamental interests of the two countries and the two peoples. The common interests of the two countries in a wide range of areas are an endogenous driving force for the sustainable development of their relations. The strategic pressure of the West is also an external factor in the development of Sino-Russian cooperation.

The framework of cooperation of the 'Silk Road Economic Belt', which China stands for, does not have a fundamental conflict of interests with the construction of the EAEU. On the contrary, they can be useful to each other. In the long term, successful coordination of the two projects can effectively eliminate the serious discrepancy between political development and economic cooperation in Sino-Russian relations. This is a strategic step to create a Sino-Russian community of interests and strengthen the Sino-Russian comprehensive strategic partnership.